\documentclass[transmag]{IEEEtran}
\usepackage{latexsym}
\usepackage[T1]{fontenc}
\def\BibTeX{{\rm B\kern-.05em{\sc i\kern-.025em b}\kern-.08em T\kern-.1667em\lower.7ex\hbox{E}\kern-.125emX}}

\usepackage{amssymb,amsmath,amsfonts,amsthm}
\usepackage{mathrsfs}
\usepackage{dsfont}
\usepackage{cite}
\usepackage{array}
\usepackage{tabularx}
\usepackage[table]{xcolor}
\usepackage{color}
\usepackage{mathtools}
\usepackage{graphicx}
\usepackage{caption}
\usepackage{subcaption}
\usepackage{mathtools}
\usepackage{tikz,pgf}
\usetikzlibrary{calc,positioning,mindmap,trees,decorations.pathreplacing}
\usepackage{bbm}
\usepackage[utf8]{inputenc}
\usepackage{algorithm}

\newcommand\irregularline[2]{%
	let \n1 = {rand*(#1)} in
	+(0,\n1)
	\foreach \a in {0.1,0.125,...,#2}{
		let \n1 = {rand*(#1)} in
		-- +(\a,\n1)
	} 
}

\def\BibTeX{{\rm B\kern-.05em{\sc i\kern-.025em b}\kern-.08em T\kern-.1667em\lower.7ex\hbox{E}\kern-.125emX}}

\newtheorem{theorem}{Theorem}
\newtheorem{lemma}{Lemma}

\newtheorem{remark}{Remark}

\newcommand{\maximize}{\mathop{\rm maximize}\limits}
\newcommand{\minimize}{\mathop{\rm minimize}\limits}

\begin{document}
	\title{Optimizing Information Freshness in a Multiple Access Channel with Heterogeneous Devices}
	
	\author{Zheng~Chen, Nikolaos~Pappas, Emil~Bj\"{o}rnson, and Erik~G.~Larsson
		\thanks{Z.~Chen, E.~Bj\"{o}rnson, and E.~G.~Larsson are with the Department of Electrical Engineering, Link\"{o}ping University,  58183 Link\"{o}ping, Sweden. Email: \{zheng.chen, emil.bjornson, erik.g.larsson\}@liu.se.}%
		\thanks{N. Pappas is with the Department of Science and Technology, Link\"{o}ping University, 60174 Norrk\"{o}ping, Sweden. Email: nikolaos.pappas@liu.se. }
		\thanks{A part of this work was presented in IEEE INFOCOM Workshops (AoI'19) \cite{aoi_infocom}. This work was supported in part by ELLIIT, CENIIT, and the Swedish Foundation for Strategic Research (SSF).}}	
	
	\IEEEtitleabstractindextext{\begin{abstract}
			In this work, we study age-optimal scheduling with stability constraints in a multiple access channel with two heterogeneous source nodes transmitting to a common destination. The first node is connected to a power grid and it has randomly arriving data packets. Another energy harvesting (EH) sensor monitors a stochastic process and sends status updates to the destination. We formulate an optimization problem that aims at minimizing the average age of information (AoI) of the EH node subject to the queue stability condition of the grid-connected node. First, we consider a \textit{Probabilistic Random Access} (PRA) policy where both nodes make independent transmission decisions based on some fixed probability distributions. We show that with this policy, the average AoI is equal to the average peak AoI, if the EH node only sends freshly generated samples. In addition, we derive the optimal solution in closed form, which reveals some interesting properties of the considered system. Furthermore, we consider a \textit{Drift-Plus-Penalty} (DPP) policy and develop AoI-optimal and peak-AoI-optimal scheduling algorithms using the Lyapunov optimization theory. Simulation results show that the DPP policy outperforms the PRA policy in various scenarios, especially when the destination node has low multi-packet reception capabilities. \end{abstract}
		
		}
	
	\maketitle
	
	\section{Introduction}
	\IEEEPARstart{T}{he} Age of Information (AoI) is a newly emerged metric and tool to capture the timeliness and freshness of data reception \cite{kosta2017age, sunmodiano2019age, Altman2010, KaulSECON2011, KaulINFOCOM2012}.
	Consider a monitored source node which generates time-stamped status updates, and transmits them through a wireless channel or through a network to a destination. The AoI that the destination has for the source is the elapsed time since the generation of the last received update. Keeping the average AoI small corresponds to having fresh information, which is critical for time-sensitive applications in the Internet of Things (IoT) scenarios and future wireless systems \cite{zhou2019joint,6g-wireless}. This notion has been extended to other metrics such as the value of information, cost of update delay, and non-linear AoI \cite{Sun2018sampling,non-linear-AoI}. 
	
	The deployment of energy harvesting (EH) sensors is envisioned as an efficient solution for energy-efficient and self-sustainable networks, especially in the IoT networks where devices opportunistically transmit small amounts of data with low power consumption \cite{Abd-Elmagid2018}. Sensors with EH capabilities can convert ambient energy (e.g., solar power or thermal energy) into electrical energy that can be used for sending status updates to the destination nodes.
	
	Though AoI analysis and optimization have been investigated in many different systems and scenarios, prior works in the literature consider either single access channel or multiple access channel (MAC) with orthogonal scheduling. In our work, we expand the current literature to the case with a MAC channel and multi-packet reception capabilities at the receiver side. Studying age-optimal status updating with EH nodes, especially with the presence of interference in a MAC channel, is a non-trivial task.

	\subsection{Related Works}
	In \cite{BacinoglouITA2015}, the authors consider the problem of optimizing the process of sending updates from an EH source to minimize the time average age of updates. Similar analysis can be found in \cite{lazy_timely, ArafaAsilomar2017, WuTGCN2018, age_eh3, ArafaICC2018, ArafaITA2018, age_eh1,alarms}.
	In \cite{BakninaCISS2018}, an erasure channel is considered where an EH-enabled transmitter sends coded status updates to the receiver to minimize the AoI. In \cite{BakninaISIT2018}, an EH transmitter is assumed to encode a message into the timings of the status updates.
	The age-energy tradeoff is explored in \cite{BacinogluJCN2019}, where a finite-battery source is charged intermittently by Poisson energy arrivals. The timeliness-distortion tradeoff of an EH-powered system is investigated in \cite{iot-eh}. In \cite{FengINFOCOMWKSHPS2018}, the optimal status updating policy for an EH source with a noisy channel is investigated. The possibility of update failures is considered in \cite{FengISIT2018}. 
	
	In addition to the case with nodes harvesting ambient energy, some other works have considered wireless power transfer (WPT) to convert the received radio frequency signals to electric power. 
	In \cite{Krikidis2018}, the performance of a WPT-powered sensor network in terms of the average AoI was studied. The work in \cite{TCOM2020-AbdElmagid} considers freshness-aware IoT networks with EH-enabled IoT devices. More specifically, the optimal sampling policy for IoT devices that minimizes the long-term weighted sum-AoI is investigated.

	In a network with multiple source nodes, assuming multi-packet reception (MPR) capabilities at the receiver, the transmissions from multiple source nodes will be successful with some probabilities that depend on the received signal-to-interference-plus-noise ratio (SINR) \cite{AlohaVerdu,Naware}. Recently, some works have considered different types of traffic associated with different source nodes, e.g., some nodes generate time-sensitive status updates, and other nodes strive to achieve as large throughput as possible. The impact of heterogeneous traffic on the AoI and the optimal update policy has been investigated in \cite{Stamatakis2018}. The work in \cite{game_coexistence} investigates Nash and Stackelberg equilibrium strategies for DSRC and WiFi coexisting networks, where DSRC and WiFi nodes are age-oriented and throughput-oriented, respectively.

	In \cite{LengTCCN2019}, dynamic programming based on a Markov Decision Process is applied in a cognitive radio network with an EH secondary user opportunistically transmitting status updates to its destination.  
	Age-optimal scheduling policies in a network with general interference constraints are studied in \cite{Talak}. In \cite{optimal-age-thpt}, several policies are considered to solve a weighted AoI minimization problem with throughput constraints. In \cite{fountoulakis2020optimal}, the sampling cost is taken into consideration in an age-optimal sampling and scheduling problem. The Drift-Plus-Penalty (DPP) policy considered in \cite{Talak, optimal-age-thpt, fountoulakis2020optimal} is developed from the Lyapunov optimization theory \cite{georgiadis2006resource, neely2010stochastic}, which is often used for solving stochastic network optimization problems with stability constraints.

	\subsection{Contributions}
	We consider a time-slotted MAC where one grid-connected user has random data arrivals and one EH sensor sends status updates to a common destination. This model is the smallest non-trivial instance of a heterogeneous network with various types of devices and different performance characteristics. 
	The main contributions of this works are summarized as follows.
	\begin{enumerate}
		\item We formulate an optimization problem that jointly considers the age minimization of the EH node with the queue stability of the grid-connected node.
		Two approaches are applied to solve this problem, namely the probabilistic random access (PRA) policy and the DPP policy.
		\item We show that with the PRA policy, the average AoI of the EH node is equal to the average peak AoI (or PAoI), which is inversely proportional to the throughput of the EH node. The optimal transmit probabilities are derived in closed form. 
		\item Simulation results show that the DPP policy clearly outperforms the PRA policy by achieving lower average AoI and PAoI, especially when the destination node has low MPR capabilities, which is an expected result. Another interesting observation is that with the DPP policy, minimizing the AoI does not give the same solution as minimizing the PAoI, while with the PRA policy, these two optimization problems are equivalent.
	\end{enumerate}
	
	Compared to the conference version in \cite{aoi_infocom}, which studies only the average AoI with the PRA policy, in this journal version, we consider both average AoI and PAoI optimization problems and extend our analysis by investigating age-optimal scheduling with both PRA and DPP policies. 
	
	\section{System Model}
	
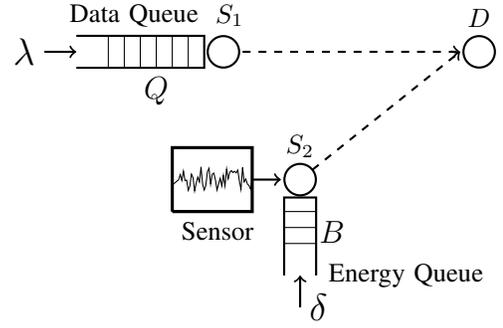
\begin{figure}[t!]
		\centering
		\begin{tikzpicture}[scale=0.85]
			\draw[thick, ->] (0,5) -- (0.5,5) ;
			\node[left] at (0,5) {\Large $\lambda$};
			\draw	(1.4,5.9) node[anchor=north] {Data Queue};
			\draw[thick] (0.5,5.25)--(2.5,5.25)-- (2.5,4.75)-- (0.5, 4.75);
			\node[below] at (1.75,4.75) {\large $Q$};
			\draw[-] (2.25,5.25)--(2.25,4.75);
			\draw[-] (2,5.25)--(2,4.75);\draw[-] (1.75,5.25)--(1.75,4.75); \draw[-] (1.5,5.25)--(1.5,4.75); \draw[-] (1.25,5.25)--(1.25,4.75); \draw[-] (1,5.25)--(1,4.75);
			\draw [thick,fill=white](2.8,5)circle[radius= 0.7 em]  ;
			\node[above] at (2.9,5.25) {$S_1$};
			\draw[thick, dashed,->] (3.1,5)--(6.5,5);
			\draw [thick,fill=white](6.8,5)circle[radius= 0.7 em]  ;
			\node[above] at (6.8,5.25) {$D$};
			\draw[thick, dashed,->] (4,3)--(6.5,5);
			\draw [thick,fill=white](4,3)circle[radius= 0.7 em]  ;
			\node[above] at (4,3.2) {$S_2$};
			\draw[thick] (3.75,1.5)--(3.75,2.72)-- (4.25,2.72)-- (4.25, 1.5);
			\draw[-] (3.75,2.5)--(4.25,2.5); 	\draw[-] (3.75,2.25)--(4.25,2.25); 	\draw[-] (3.75,2)--(4.25,2);
			\draw[thick, ->] (4,1) -- (4,1.5) ;
			\node[right] at (4,1) {\Large $\delta$};
			\node[below] at (4.5,2.5) {\large $B$};
			\node[right] at (4.3,1.5) {Energy Queue};
			\draw[thick, ->] (3.25,3) -- (3.7,3) ;
			\draw[very thick] (2,2.5)--(2,3.5)-- (3.25,3.5)-- (3.25, 2.5) --(2,2.5);
			\node[right] at (2,2.2) {Sensor};
			\draw[black] (2,3) \irregularline{0.2cm}{1.25};
		\end{tikzpicture}
		\caption{The system model. One grid-connected source node and an EH sensor share the same wireless channel to a common destination. The EH sensor is generating status updates to transmit to the destination.}
		\label{fig:system}
	\end{figure}

	We consider a time-slotted MAC where two heterogeneous source nodes intend to transmit to a common destination $D$, as shown in Fig.~\ref{fig:system}. The first node $S_1$ is connected to a power grid, thus its activities are not battery-constrained.  
	The data packets arrive at the queue of $S_1$ following a Bernoulli process with probability $\lambda$. Denote by $Q(t)$ the data queue size of node $S_1$ in time slot $t$, which has infinite capacity.\footnote{Our work can handle the case of a finite queue, by replacing the probability of the queue being empty by the new expression with the finite queue size. When considering a finite-capacity queue, the packet dropping probability is a more relevant metric than the stability, which will result in a different problem formulation with more elaborated expressions. This specific consideration will not bring much useful insights into this problem, as the general observations can still hold.} The second node $S_2$ is not connected to a dedicated power source, but it can harvest energy from the environment. We assume that the battery charging process follows a Bernoulli process with probability $\delta$, with $B(t)$ representing the number of energy units in the energy source (battery) at node $S_2$ in time slot $t$. The capacity of the battery is assumed to be infinite.\footnote{The infinite battery size is a common assumption in the literature, such as \cite{ArafaAsilomar2017, WuTGCN2018, ArafaICC2018, BakninaCISS2018, FengINFOCOMWKSHPS2018}. Considering a finite battery size will only make the expressions more elaborated without bringing more values to this problem.} 
	
	The two source nodes generate different types of data traffic which are associated with different performance goals. Node $S_1$ sends data packets that have been stored in its queue. The average delay to receive a packet from $S_1$ will be finite if the data queue is stable. Node $S_2$ always generates the freshest sample of the status update and sends it to the destination when it decides to transmit, i.e., the status update transmitted at time slot $t$ is generated right before the transmission and we assume that the sampling is instantaneous. We consider equal-sized data packets and the transmission of one packet occupies one time slot.

	Let $\{a_1(t)\}$ and $\{a_2(t)\}$ represent the Bernoulli processes for the data arrivals at node $S_1$ and energy arrivals at node $S_2$, respectively. We have  $\lambda=\mathbb{E}[a_1(t)]$ and $\delta=\mathbb{E}[a_2(t)]$. We consider an early-departure-late-arrival model and first-come-first-served principle for the queue. 
	The data queue of node $S_1$ is updated following the equation
	\begin{equation}
		Q(t+1)=\max[Q(t)-b_1(t),0]+a_1(t),
		\label{eq:queue-update}
	\end{equation}
	where $b_1(t)=1$ if the destination successfully receives a packet from $S_1$ in time slot $t$, otherwise $b_1(t)=0$.
	In the case of an unsuccessful transmission from $S_1$, the packet has to be re-transmitted in a future time slot. We assume that the receiver gives an instantaneous error-free acknowledgment (ACK) feedback of all the packets that were successful in a slot at the end of the slot. Then, $S_1$ removes the successfully transmitted packets from its queue. 
	
	We assume that both nodes have a fixed transmit power, and for $S_2$ the transmission of one status update consumes one energy unit from the battery. The same assumption is also made in \cite{age_eh1, age_eh3, FengISIT2018}.
	The battery queue of $S_2$ evolves according to the equation
	\begin{equation}
		B(t+1)=\max[B(t)-u_2(t),0]+a_2(t),
	\end{equation}
	where $u_2(t)=1$ if node $S_2$ attempts to transmit a status update in time slot $t$, otherwise $u_2(t)=0$. 
	When an update is transmitted from $S_2$, in case of a transmission failure, the packet is dropped, and a new update will be generated for its next attempted transmission.

	\subsection{PHY Model and Success Probabilities}
	Since both source nodes are transmitting to a common destination, we assume MPR capabilities at the destination node $D$, which means that $D$ can decode packets from multiple simultaneous transmissions that are interfering with each other. The transmission from one source node is successful if its received signal-to-interference-plus-noise ratio (SINR) at $D$ exceeds a certain threshold \cite{AlohaVerdu}.
	
	We assume a block fading channel, where the fading coefficient remains constant during one time slot and independently changes from one slot to another. 
	The received SINR at node $i$ is given by
	\begin{equation*}
		{\rm SINR}_{i}=\frac{|h_{i}|^2\tilde{\beta}_i}{\sum_{j\in \mathcal{A}\backslash\left\{i\right\}} |h_{j}|^2 \tilde{\beta}_j+1},
	\end{equation*}
	where $\mathcal{A}$ denotes the set of active transmitters; $h_{i}$ denotes the small-scale channel fading of node $i$, which follows i.i.d. $\mathcal{CN}(0,1)$ distribution; $\tilde{\beta}_i$ denotes the large-scale fading of link $i$, normalized over the transmit power and noise variance.

	We define $p_{i/i}$ as the success probability of $S_i,~i\in\{1,2\}$, when only $S_i$ is transmitting, and $p_{i/i,j}$ as the success probability of $S_i$ when both $S_i$ and $S_j$ are transmitting and interfering with each other. Denote by $\theta_i$ the SNR/SINR threshold for successful transmission. Using the small-scale fading distribution, we have
	\begin{equation}
		p_{i/i}=\mathbb{P} \left\lbrace \mathrm{SNR}_{i} \geq \theta_i \right\rbrace = \exp \left(- \theta_i /\tilde{\beta}_i\right),\text{ }i=\{1,2\}, \label{eq:suc-prob1} 
	\end{equation}
	\vspace{-0.3cm}
	\begin{equation} 
		p_{i/i,j}=\mathbb{P} \left\lbrace \mathrm{SINR}_{i} \geq \theta_i \right\rbrace = \frac {\exp \left(- \theta_i/\tilde{\beta}_i\right)}{ 1+\theta_i \tilde{\beta}_j/\tilde{\beta}_i} , i=\{1,2\}, j\neq i.\label{eq:suc-prob2}  
	\end{equation}
	
	Note that our analysis holds for any specific channel model, and the analysis on the AoI and stability only requires the success probabilities.

	\subsection{Age of Information}
	\begin{figure}
		\centering
		\begin{tikzpicture}[scale=0.95]
			\draw[->] (0,0) -- (8.2,0) node[anchor=north] {$t$};
			\draw[->] (0,0) -- (0,3.5) node[anchor=east] {$A(t)$};
			
			\draw	(-0.3,0.25) node[anchor=south] {$1$};
			\draw	(-0.3,0.7) node[anchor=south] {$2$};
			\draw[thick]  (-0.1,0.5) -- (0.1,0.5); 
			\draw[thick]  (-0.1,1) -- (0.1,1); 
			\draw[thick]  (-0.1,1.5) -- (0.1,1.5); 
			\draw[thick]  (-0.1,2) -- (0.1,2); 
			\draw[thick]  (-0.1,2.5) -- (0.1,2.5); 
			\draw[fill=gray!10] (1,0)--(1,0.5)-- (1.5,0.5)-- (1.5,1)-- (2,1)-- (2,1.5)-- (2.5,1.5)-- (2.5,2)-- (3,2)--(3,0);
			\draw[fill=gray!10] (4,0) --(4,0.5) -- (4.5,0.5) -- (4.5,1)-- (5,1)-- (5,1.5)-- (5.5,1.5)-- (5.5,2)-- (6,2)-- (6,2.5)-- (6.5,2.5)-- (6.5,3)--(7,3)--(7,0);

			\draw[->,>=stealth]    (1,0) -- (1,-0.4) node[anchor=south,below] {$t_1$};
			\draw[->,>=stealth]  (3,0) -- (3,-0.4) node[anchor=south,below] {$t_2$};
			\draw[->,>=stealth]  (4,0) -- (4,-0.4) node[anchor=south,below] {$t_k$};
			\draw[->,>=stealth]   (7,0) -- (7,-0.4) node[anchor=south,below] {$t_{k+1}$};
			\draw[<-] (2,0.5) to [out=95,in=250] (1.5,1.5) node [above] {{$Y_1$}};   
			\draw[<-] (5.5,1) to [out=95,in=250] (5,2) node [above] {{$Y_k$}};   
			
			\draw [<->] (4,0.25) -- (5,0.25) node[pos=.75,sloped,above] {$T_1$} ;
			\draw[thick]  (4,0.15) -- (4,0.35);    
			\draw [<->] (6,0.25) -- (7,0.25) node[pos=.5,sloped,above] {$T_{M}$} ;
			\draw[thick]  (5,0.15) -- (5,0.35);      
			\draw[thick]  (6,0.15) -- (6,0.35);    
			\draw[dotted]  (5,0.25) -- (6,0.25);    
			\draw[thick]  (7,0.15) -- (7,0.35);   
			
			\draw [<->] (1,-1) -- (3,-1) node[pos=.5,sloped,below] {$X_1$} ;
			\draw[thick]  (1,-1.1) -- (1,-0.9) 
			(3,-1.1) -- (3,-0.9);
			
			\draw [<->] (4,-1) -- (7,-1) node[pos=.5,sloped,below] {$X_k$} ;
			\draw[thick]  (4,-1.1) -- (4,-0.9); 
			\draw[thick]  (7,-1.1) -- (7,-0.9) ;
			
			\draw[thick] (0,0.5) -- (0.5,0.5) -- (0.5,1)-- (1,1)-- (1,0.5)-- (1.5,0.5)-- (1.5,1)-- (2,1)-- (2,1.5)-- (2.5,1.5)-- (2.5,2)-- (3,2)--(3,0.5)--(3.5,0.5);
			\draw[white, fill=white!50] (3.5,-0.2) -- (3.5,0.2) -- (3.95,0.2) -- (3.95,-0.2) ;   
			\draw[dashed]  (3.5,0) -- (4,0); 
			\draw[thick] (4,0.5) -- (4.5,0.5) -- (4.5,1)-- (5,1)-- (5,1.5)-- (5.5,1.5)-- (5.5,2)-- (6,2)-- (6,2.5)-- (6.5,2.5)-- (6.5,3)--(7,3)--(7,0.5)--(7.5,0.5);
			
			\draw [decorate,decoration={brace,amplitude=5pt,mirror,raise=2pt},yshift=0pt] (3.1,0.05) -- (3.1,2) node [black,midway,yshift=0.1cm,xshift=0.5cm] {$X_1$};
			\draw [decorate,decoration={brace,amplitude=5pt,mirror,raise=2pt},yshift=0pt] (7.1,0.05) -- (7.1,3) node [black,midway,xshift=0.5cm] { $X_k$};
		\end{tikzpicture}
		\vspace{-0.4cm}
		\caption{Evolution of the AoI. $t_k$ denotes the time when the destination received the $k$-th update. $Y_k$ is the total area below the AoI step line between $t_k$ and $t_{k+1}$. $X_k$ is the number of time slots between the successful receptions of the $k$-th and the $(k+1)$-th status updates.}
		\label{fig:aoi}
		\vspace{-0.4cm}
	\end{figure}
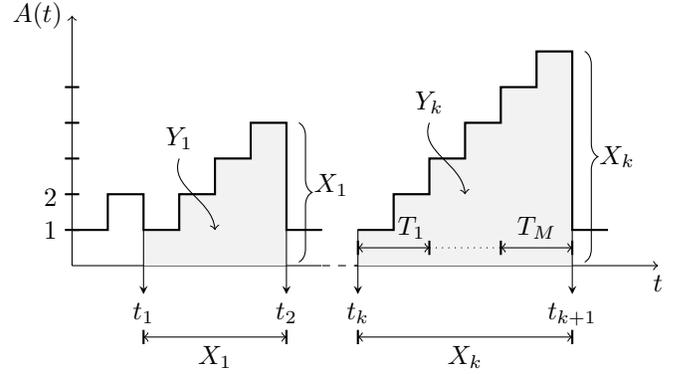
	
	At time slot $t$, the AoI seen at the destination is defined by the difference between the current time $t$ and the time slot $G(t)$ when the latest successfully received update was generated. Following the standard definition \cite{kosta2017age,sunmodiano2019age}, the AoI at slot $t$ is given by 
	\begin{equation}
		A(t)=t-G(t).
	\end{equation}
	As we consider slotted transmissions, the AoI takes integer values, i.e., $A(t)\in\{1,2,\ldots\}$, as shown in Fig.~\ref{fig:aoi}. Since each transmitted update by node $S_2$ is always generated at the end of the previous slot, the AoI drops to $1$ when there is a successful reception of a status update at the destination. The evolution of the AoI between two consecutive slots can be written as
	\begin{equation}
		A(t+1)= 
		\left\lbrace 
		\begin{array}{ccc}
			1
			& \text{if successful reception at slot $t$}, \\
			A(t)+1
			& \text{otherwise}.
		\end{array} \right.
		\label{eq:age_definition}
	\end{equation}
	The average AoI is defined as
	\begin{equation}
		\overline{A}=\limsup\limits_{t\rightarrow\infty}\mathbb{E}\left[\frac{1}{t}\sum_{\tau=0}^{t-1}A(\tau)\right].
	\end{equation}
	
	Upon each successful reception of a status update, the value of the AoI before dropping to $1$ is counted as one peak. Let $T_s(t)=\mathds{1}\{\text{successful~reception~at~slot}~t\}$ be the process representing the transmission success/failure in each slot. Then \eqref{eq:age_definition} can be written as
	\begin{equation}
		A(t+1)=A(t)+1-T_s(t) A(t).
	\end{equation}
	When $T_s(t)=1$, there is a peak value at slot $t$. 
	The average PAoI is defined by
	\begin{equation}
		\overline{A}_p=\limsup\limits_{t\rightarrow\infty} \frac{\mathbb{E}\left[\sum_{\tau=0}^{t-1}T_s(\tau) A(\tau)\right]}{\mathbb{E}\left[\sum_{\tau=0}^{t-1}T_s(\tau) \right]}.
		\label{eq:peak_age_def}
	\end{equation}

	\subsection{Problem Formulation}
	Since both source nodes share the same wireless channel, scheduling plays an important role. We define the scheduling policy space by $\Pi=\left\{u_i(t)\in\{0,1 \},\forall t, i=\{1,2\}\right\}$, where $u_i(t)$ is the scheduling decision of node $i$ at slot $t$. $u_i(t)=1$ means that node $S_i$ is scheduled to transmit in time slot $t$, and $u_i(t)=0$ means that node $S_i$ is inactive. 
	
	We aim at finding an optimal scheduling policy $\pi\in\Pi$ that minimizes the average age of node $S_2$, while keeping the data queue of $S_1$ stable. We consider both average AoI and average PAoI as the age-performance metric.
	The optimization problem is defined by
	\begin{subequations}
		\begin{align}
			\minimize\limits_{ \pi\in \Pi} ~~&\overline{A} \\
			\textrm{subject~to}~~& \overline{Q}< \infty, \label{eq:constraint}
		\end{align}
	\end{subequations}
	where $\overline{Q}=\limsup\limits_{t\rightarrow\infty}\mathbb{E}\left[\frac{1}{t}\sum_{\tau=0}^{t-1}Q
	(\tau)\right]$ is the time-average expectation of the queue size.
	The constraint in \eqref{eq:constraint} ensures that the data queue of $S_1$ is strongly stable. This problem is feasible if and only if $\lambda<p_{1/1}$.\footnote{When $\lambda<p_{1/1}$, we can always find a scheduling policy that satisfies the queue stability condition by reducing the transmissions of the EH node.}
	
	The stability condition of node $S_1$ depends on its service probability. The activity and battery status of node $S_2$ can affect the interference that it causes to $S_1$. In such way, the AoI of node $S_2$ and the queue stability of node $S_2$ are coupled. 
	Due to the random energy arrivals and the interference between the two users, this age-optimal scheduling problem is more difficult to solve than in the case with orthogonal scheduling and without EH.
	
	To solve this problem, we first consider a random access policy, where the transmission actions of both nodes follow some fixed probability distributions, and they are independent of each other. 
	Another option is to consider the DPP policy, where the scheduling decision in each time slot is based on the observed network state in that slot, such as the queue size and battery size.
	The details of these two policies will be presented in Sections~\ref{sec:stationary} and \ref{sec:dynamic}.

	\section{Probabilistic Random Access Policy}
	\label{sec:stationary}
	First, we consider a decentralized probabilistic policy with random transmission decisions at both nodes following some stationary probability distributions. 
	The PRA policy is described as follows:
	\begin{itemize}
		\item When the data queue of $S_1$ is not empty, it transmits a packet to the destination with probability $q_1$. 
		\item When $S_2$ has a non-empty battery, it generates a status update with probability $q_2$ and transmits it to the destination.
	\end{itemize}
	With this policy, each node makes independent decisions without coordinating with each other.\footnote{Note that the considered PRA policy serves as a baseline of stationary randomized scheduling policies, and it is not guaranteed to be the best policy in this category. }
	To solve the age minimization problem with stability constraint, we first characterize the stability condition of node $S_1$. Then, we derive the average AoI and average PAoI of node $S_2$, which are given as functions of the probabilities $q_1$, $q_2$, $\delta$, and $\lambda$. 
	
	\subsection{Stability Analysis of Node $S_1$}
	\label{sec:primary}
	The service probability of $S_1$ represents the probability of transmission success when $S_1$ attempts to transmit a packet. It can be obtained by averaging the success probabilities over three cases: $S_2$ has empty battery, $S_2$ has non-empty battery but decides not to transmit, and $S_2$ decides to transmit. The service probability of $S_1$ is 
	\begin{align}
		\mu=&~\mathbb{P}\left[B(t)=0\right]q_1 p_{1/1}+\mathbb{P}\left[B(t)\neq 0\right]q_1 (1-q_2)p_{1/1} \nonumber\\&+\mathbb{P}\left[B(t)\neq 0\right]q_1 q_2p_{1/1,2}\nonumber\\
		=&~q_1p_{1/1}\left(1-q_2\mathbb{P}\left[B(t)\neq 0\right]\right)+q_1\mathbb{P}\left[B(t)\neq 0\right]q_2p_{1/1,2}.
		\label{eq:mu}
	\end{align}
	The queue of $S_1$ is stable if and only if $\lambda<\mu$. Stability implies that the queueing delay will be finite. 
	When the queue at $S_1$ is stable, $Q(t)$ has a unique stationary distribution. The probability of a non-empty queue is 
	\begin{equation}
		\mathbb{P}[Q(t)\neq 0]=\frac{\lambda}{\mu}.
		\label{eq:proba-Qnonzero}
	\end{equation}
	This probability will be used in the average AoI and PAoI analysis for node $S_2$.
	
	\subsubsection{When $S_2$ relies on EH}
	Recall that the energy arrival process at the EH node $S_2$ follows a Bernoulli process with probability $\delta$. The evolution of the energy queue $B(t)$ can be modeled as a Discrete Time Markov Chain, and it has a unique stationary distribution when the energy queue is stable. When $\delta<q_2$,  we have 
	\begin{equation}
		\mathbb{P}[B(t)\neq 0]=\frac{\delta}{q_2}.
		\label{eq:proba-nonempty}
	\end{equation} 
	Plugging \eqref{eq:proba-nonempty} into \eqref{eq:mu}, we obtain 
	\begin{equation}
		\mu=q_1 p_{1/1}(1-\delta)+q_1\delta p_{1/1,2}.
		\label{eq:mu_case1}
	\end{equation}
	The stability condition $\lambda<\mu$ yields 
	\begin{equation}
		q_1>\frac{\lambda}{p_{1/1}(1-\delta)+\delta p_{1/1,2}}.
		\label{eq:stability_q1}
	\end{equation} 
	From \eqref{eq:stability_q1}, we see that in order to have a stable queue at $S_1$, the transmit probability $q_1$ needs to be higher than a threshold, and this threshold is independent of $q_2$ when $\delta<q_2$. This is because when $\delta<q_2$, how frequently $S_2$ is causing interference to the transmission of $S_1$ is only limited by the energy arrival probability. 
	
	Note that when $\lambda\geq p_{1/1}(1-\delta)+\delta p_{1/1,2}$, we cannot find any probability $q_1\in[0,1]$ that satisfies the stability condition in \eqref{eq:stability_q1}, if we keep the assumption of $\delta<q_2$. Therefore, we consider the case when $\delta\geq q_2$. In this case, the energy queue is unstable because its associated Markov chain is not positive recurrent, thus the energy queue size does not have a unique stationary distribution.\footnote{When the energy queue does not have a unique stationary distribution, asymptotically the probability of returning to the zero state (empty energy queue) is zero, i.e, $\mathbb{P}[B=0]=0$. Therefore, from the practical perspective, the battery will always be charged in the long term if the energy queue is unstable.}
	Then, we can disregard the energy queue and consider the system as if $S_2$ were connected to a power grid. In the remainder of this section, we always divide our analysis into two parts: (1) when $S_2$ relies on EH and $\delta<q_2$; (2) when $S_2$ is connected to a power grid.
	
	\subsubsection{When $S_2$ is connected to a power grid}
	In this case, $\mathbb{P}[B(t)\neq 0]=1$. From \eqref{eq:mu}, we have 
	\begin{equation}
		\mu=q_1 p_{1/1}(1-q_2)+q_1 q_2p_{1/1,2}.
		\label{eq:mu_case2}
	\end{equation}
	The queue is stable if and only if 
	\begin{equation}
		q_1>\frac{\lambda}{p_{1/1}(1-q_2)+q_2p_{1/1,2}}.
		\label{eq:q1_connected}
	\end{equation}
	From the above inequality, if $p_{1/1}>\lambda$ and $p_{1/1}>p_{1/1,2}$ holds, we can always find $q_1$ and $q_2$ that satisfy the stability condition of $S_1$. 
	
	Combining the two cases, we obtain the following lemma.
	\begin{lemma}
		\label{lemma-stability}
		The service probability of node $S_1$ is given by
		\begin{equation}
			\mu=q_1 p_{1/1}\left(1-\min\{\delta, q_2\}\right)+q_1 p_{1/1,2} \cdot \min\{\delta, q_2\}.
			\label{eq:mu_general}
		\end{equation}
		The queue at $S_1$ is stable if and only if 
		\begin{equation}
			q_1>\frac{\lambda}{p_{1/1}-\min\{\delta, q_2\}\cdot \left(p_{1/1}-p_{1/1,2}\right)}.
			\label{eq:stability_general}
		\end{equation}
	\end{lemma}

	\subsection{Average AoI of Node $S_2$}
	For a period of $N$ time slots where $K$ successful updates occur, from Fig.~\ref{fig:aoi}, the average AoI can be computed as 
	\begin{equation}
		A_N=\frac{1}{N}\sum\limits_{k=1}^{K}Y_k=\frac{K}{N}\frac{1}{K}\sum\limits_{k=1}^{K}Y_k.
	\end{equation}
	Let $X$ be a random variable (RV) denoting the time difference between two successful receptions of status updates. We have $\lim\limits_{N\rightarrow\infty}\frac{K}{N}=\frac{1}{\mathbb{E}[X]}$. Meanwhile, $\frac{1}{K}\sum_{k=1}^{K}Y_k$ is the arithmetic mean of $Y$, which converges almost surely to $\mathbb{E}[Y]$ when $K\rightarrow\infty$. Then we have the average AoI as
	\begin{equation}
		\overline{A}=\lim\limits_{N\rightarrow\infty}A_N=\frac{\mathbb{E}[Y]}{\mathbb{E}[X]}.
	\end{equation}
	From Fig.~\ref{fig:aoi}, we observe the relation between $Y_k$ and $X_k$ as follows
	\begin{equation}
		Y_k=\sum\limits_{m=1}^{X_k}m=\frac{1}{2}X_{k}(X_{k}+1).
	\end{equation}
	Then, we obtain
	\begin{equation}
		\overline{A}=\frac{\mathbb{E}\left[\frac{X_k^2}{2}+\frac{X_k}{2}\right]}{\mathbb{E}[X]}=\frac{\mathbb{E}[X^2]}{2\mathbb{E}[X]}+\frac{1}{2}.
		\label{eq:aoi}
	\end{equation}
	The average AoI with the PRA policy is given as follows.
	\begin{theorem}\label{theorem1}
		With the PRA policy, when the queue at node $S_1$ is stable, i.e., $\lambda<\mu$, the average AoI of node $S_2$ is
		\begin{equation}
			\overline{A}=\frac{1}{\overline{p}_2\cdot \min\{\delta,q_2\}},
			\label{eq:average-AoI}
		\end{equation}
		where $\overline{p}_2$ is the success probability of node $S_2$, given as
		\begin{equation}
			\overline{p}_2=p_{2/2}- \frac{(p_{2/2}-p_{2/1,2}) \lambda}{p_{1/1}-\min\{\delta, q_2\}\cdot \left(p_{1/1}-p_{1/1,2}\right)}.
		\end{equation}
	\end{theorem}
	\begin{IEEEproof}
		See Appendix \ref{appen_theorem}.
	\end{IEEEproof}
	
	\begin{remark}
		When $\delta<q_2$, the average AoI is limited by the energy arrival probability $\delta$. Otherwise, if $S_2$ is constantly charged, the average AoI depends on the energy departure probability $q_2$. The average AoI of the EH node is inversely proportional to its throughput.
	\end{remark}

	\subsection{Optimization Problem}
	The average AoI optimization problem is defined by
	\begin{subequations}
		\begin{align}
			\minimize\limits_{ q_1,q_2} ~~&\frac{1}{\overline{p}_2\cdot \min\{\delta,q_2\}}\\
			\textrm{subject~to}~~& \mu>\lambda,\\
			& q_1, q_2\in[0,1],
		\end{align}
		\label{eq:optim_general}%
	\end{subequations}
	where the service probability $\mu$ is given in \eqref{eq:mu_general}. The problem is feasible if and only if $\lambda<p_{1,1}$.
	Depending on the data and energy arrival rates, the stability condition leads to different sub-cases where the optimal values of $q_1$ and $q_2$ are given in the following lemma.
	
	\begin{lemma}
		\label{lemma_optimal_proba}
		The optimal transmit probabilities can be obtained in the following two cases:
		\begin{itemize}
			\item If $0<\delta< \min\left\{\frac{p_{1/1}-\lambda}{p_{1/1}-p_{1/1,2}},1\right\}$, there exists $\delta<q_2\leq1$ that satisfies the queue stability condition. The optimal solution is 
			\begin{align}
				q_1^*&>\frac{\lambda}{p_{1/1}(1-\delta)+\delta p_{1/1,2}}, \label{eq:optimal_q1_case1}\\
				q_2^*&>\delta. \label{eq:optimal_q2_case1}
			\end{align}
			\item If $\delta\geq  \min\left\{\frac{p_{1/1}-\lambda}{p_{1/1}-p_{1/1,2}},1\right\}$, the stability condition implies $q_2\leq\lambda$. 
			The optimal solution is 
			\begin{align}
				q_1^*&=1, \label{eq:q1opt}\\
				q_2^*&=\min\left\{\frac{p_{1/1}-\lambda}{p_{1/1}-p_{1/1,2}},\delta\right\}-\xi, \label{eq:q2opt}
			\end{align}
			where $\xi>0$ is a sufficiently small positive value to ensure that the service probability of node $S_1$ is strictly larger than the arrival probability.
		\end{itemize}
	\end{lemma}
	\begin{IEEEproof}
		See Appendix \ref{appen1}.
	\end{IEEEproof}
	
	\begin{remark}
		Since $q_1^*=1$ is a special case of the optimal solution given in \eqref{eq:optimal_q1_case1}, the minimum AoI of node $S_2$ is always achieved by letting node $S_1$ transmit with probability $1$, and choosing the largest $q_2$ that guarantees the stability of $S_1$.
	\end{remark}
	
	\subsection{Average PAoI of Node $S_2$}
	\label{sec:paoi}
	From Fig.~\ref{fig:aoi}, it is straightforward to establish the relation
	\begin{align}
		\overline{A}_p=&~\mathbb{E}[X]=\frac{\mathbb{E}[T]}{\overline{p}_2},\label{eq:peak-aoi}
	\end{align}
	which follows from \eqref{EX-gen}. 
	After substituting \eqref{eq:ET_both} into \eqref{eq:peak-aoi}, we obtain
	\begin{equation}
		\overline{A}_p=\frac{1}{\overline{p}_2\cdot \min\{\delta,q_2\}},
		\label{eq:average-PAoI}
	\end{equation}
	which is the same as the average AoI given in \eqref{eq:average-AoI}. 
	
	Summarizing our results in this section, we have the same finding as in Lemma $1$ in \cite{Talak}.

	\begin{remark}
		\label{theom2}
		In a MAC, when a source node (with EH or not) always generates a fresh sample of the status update before transmitting to the destination, with the PRA policy, the average AoI is the same as the average PAoI, which is inversely proportional to the throughput (average number of successfully transmitted packets per slot) of this source node.
	\end{remark}

	\section{Drift-Plus-Penalty Policy}
	\label{sec:dynamic}
	In this section, we consider a DPP policy using the Lyapunov optimization framework \cite{neely2010stochastic}.
	Let $\mathbf{U}(t)=(u_1(t), u_2(t))$ represent the scheduling decision in slot $t$.  For the two-user system we consider, there are four possible scheduling decisions, i.e., $\mathbf{U}(t)\in\{(1,1), (1,0), (0,1),(0,0)\}$.
	
	In each slot $t$, depending on the scheduling decision $\mathbf{U}(t)$, the transmissions from the two source nodes will be successful with different probabilities. 
	We define the event $\mathcal{S}=\text{``transmission}$ $\text{success''}$. Then we have $b_i(t)=\mathds{1}\{\mathcal{S} | \mathbf{U}(t)\}$ as the successful transmission process of user $i$ in slot $t$ given the scheduling decision $\mathbf{U}(t)$, and 
	$p_i(t)=\mathbb{P}[\mathcal{S}|\mathbf{U}(t)]$ as the conditional success probability.
	Then we have
	\begin{equation}
		\label{eq:p1}
		p_1(t)= 
		\left\lbrace 
		\begin{array}{ccc}
			p_{1/1,2} &\text{if}~~\mathbf{U}(t)=(1,1), \\
			p_{1/1} &\text{if}~~\mathbf{U}(t)=(1,0), \\
			0 &\text{if}~~\mathbf{U}(t)=(0,1),\\
			0&\text{if}~~\mathbf{U}(t)=(0,0). 
		\end{array} \right.
	\end{equation}
	\begin{equation}
		\label{eq:p2}
		p_2(t)= 
		\left\lbrace 
		\begin{array}{ccc}
			p_{2/1,2} &\text{if}~~\mathbf{U}(t)=(1,1) , \\
			0&\text{if}~~\mathbf{U}(t)=(1,0), \\
			p_{2/2}  &\text{if}~~\mathbf{U}(t)=(0,1),\\
			0 &\text{if}~~\mathbf{U}(t)=(0,0). 
		\end{array} \right.
	\end{equation}
	
	Though we showed that the average AoI and PAoI are the same with the PRA policy, this might not hold for other policies. Therefore, in this section, we will solve the average AoI and PAoI optimization problems separately.

	\subsection{Average AoI Optimization}
	In order to have a successful status update reception, $S_2$ needs to have non-empty battery to be able to transmit, and the transmission needs to be successful. Let $H(t)=\mathds{1}\{B(t)>0\}$ indicate the battery status. Recall that the successful transmission process of $S_2$ is defined by $b_2(t)=\mathds{1}\{\mathcal{S}| \mathbf{U}(t)\}$. The average AoI updates as follows:
	\begin{equation}
		A(t+1)=A(t)+1-H(t)b_2(t)A(t).
	\end{equation}
	The AoI optimization problem is formulated as
	\begin{subequations}
		\label{prob:opti-aoi}
		\begin{align}
			\minimize ~~&\limsup\limits_{t\rightarrow\infty}\mathbb{E}\left[\frac{1}{t}\sum_{\tau=0}^{t-1}A(\tau+1)\right]\\
			\textrm{subject~to}~~& \overline{Q}< \infty, \\
			&\mathbf{U}(t)\in\{(1,1), (1,0), (0,1),(0,0)\}.
		\end{align}
	\end{subequations}
	
	Denote by $\boldsymbol{\Phi}(t)=[A(t), H(t), Q(t)]$ the network state at slot $t$, which consists of the AoI and battery status of node $S_2$, and the data queue size of node $S_1$. 
	We consider the quadratic Lyapunov function $\mathcal{L}(\boldsymbol{\Phi}(t))=\frac{1}{2}Q^2(t)$ and
	the following one-slot conditional Lyapunov drift
	\begin{equation}
		\Delta(t)=\mathbb{E}[\mathcal{L}(\boldsymbol{\Phi}(t+1))-\mathcal{L}(\boldsymbol{\Phi}(t))|\boldsymbol{\Phi}(t)].
	\end{equation} 
	We consider an AoI-related penalty function 
	\begin{equation}
		P(t)=V\mathbb{E}[A(t+1) | \boldsymbol{\Phi}(t)],
	\end{equation}
	where $V$ is a constant parameter that determines the weight on the penalty function.
	
	\begin{lemma}
		\label{lemma2}
		The DPP function $\Delta(t)+ P(t)$ is upper bounded by
		\begin{equation}
			\begin{split}
				\Delta(t)+ P(t)\leq ~&\frac{\lambda^2+1}{2}+\lambda Q(t)+V\left[A(t)+1\right]\\
				&-\mathbb{E}[p_1(t)Q(t)+V p_2(t) H(t) A(t)|\boldsymbol{\Phi}(t)].
			\end{split}
			\label{eq:dpp-bound}
		\end{equation}
	\end{lemma}
	\begin{IEEEproof}
		See Appendix \ref{appen2}.
	\end{IEEEproof}
	
	We intend to greedily minimize the DPP upper bound, by opportunistically minimizing the term inside the conditional expectation in every slot. This means that in each slot $t$, we check for all possible values of the scheduling decision $\mathbf{U}(t)$, and choose the one that maximizes $p_1(t)Q(t)+p_2(t) H(t) A(t)$.

	\begin{algorithm}[t!]
		\caption{DPP Algorithm for AoI Minimization} \label{algorithm1}
		\begin{enumerate}
			\item Initialization: $Q(0)=0$, $A(0)=0$ and $B(0)=0$. Set $t=1$ and choose appropriate value for $V$.
			\item At slot $t$, the network scheduler observes $Q(t)$, $A(t)$ and $H(t)$, and makes scheduling decision $\mathbf{U}(t)$ by solving
			\begin{equation}
				\maximize\limits_{\mathbf{U}(t)}~~ p_1(t)Q(t)+V p_2(t) H(t) A(t),
				\label{eq:solution1} \nonumber
			\end{equation} 
			where $p_1(t)$ and $p_2(t)$ are given in \eqref{eq:p1} and \eqref{eq:p2}.
			\item Update the data queue $Q(t)$, the energy queue $B(t)$, and the AoI $A(t)$ as
			\begin{align} \notag
				Q(t+1)&=\max[Q(t)-b_1(t),0]+a_1(t),\\ \notag
				B(t+1)&=\max[B(t)-u_2(t),0]+a_2(t),\\ \notag
				A(t+1)&=A(t)+1-H(t)b_2(t)A(t),
			\end{align}
			where $b_i(t)=\mathds{1}\{\mathcal{S}|\mathbf{U}(t)\}$ and $H(t)=\mathds{1}\{B(t)>0\}, \forall i=1,2$.
			\item Repeat steps 2--3 for the next slot $t+1$.
		\end{enumerate}
	\end{algorithm}
	The details of the DPP algorithm for AoI optimization are presented in Algorithm \ref{algorithm1}.
	In every slot, the network nodes report their local status information (queue size, battery size, AoI) to a centralized network scheduler. Then, the scheduling decisions are computed at the scheduler and communicated back to the nodes.

	\subsection{Average PAoI Optimization}
	From the definition of the average PAoI in \eqref{eq:peak_age_def} and $T_s(t)=H(t)\cdot b_2(t)$, we have
	\begin{equation}
		\overline{A}_p=\limsup\limits_{t\rightarrow\infty} \frac{\mathbb{E}\left[\sum_{\tau=0}^{t-1}H(\tau)b_2(\tau) A(\tau)\right]}{\mathbb{E}\left[\sum_{\tau=0}^{t-1}H(\tau)b_2(\tau) \right]}.
	\end{equation}
	It was shown in \cite{Talak} that  $\lim\limits_{t\rightarrow\infty}\mathbb{E}\left[\frac{1}{t}\sum_{\tau=0}^{t-1}H(\tau)b_2(\tau) A(\tau)\right]=1$ for any scheduling policy that guarantees bounded age. The average PAoI minimization problem $\min\limits \overline{A}_p$ is equivalent to 
	\begin{subequations}
		\label{eq:opt-trans1}
		\begin{align}
			\maximize\limits_{x>0}~~&x\label{eq:optimization_utility}\\
			\textrm{subject~to}~~&\liminf\limits_{t\rightarrow\infty} \mathbb{E}\left[\frac{1}{t}\sum_{\tau=0}^{t-1}H(\tau)b_2(\tau)\right]\geq x, \\
			& \overline{Q}< \infty,
		\end{align}
	\end{subequations}
	where $x$ is an auxiliary variable. Unlike the case with average AoI optimization in the previous section, the new problem is independent of the AoI in each time slot. Note that the solution to this optimization problem also maximizes the throughput of node $S_2$ under stability condition of node $S_1$. \textit{This suggests that with the DPP policy, throughput-optimal scheduling also minimizes the average PAoI, but not necessarily minimizes the average AoI.}
	
	We introduce a stochastic process $\alpha(t)$ which has time average $\lim\limits_{t\rightarrow\infty}\mathbb{E}\left[\frac{1}{t}\sum_{\tau=0}^{t-1}\alpha(\tau)\right]=x$. 
	The problem in \eqref{eq:opt-trans1} becomes
	\begin{subequations}
		\label{eq:opt-trans2}
		\begin{align}
			\maximize\limits_{}~~&\lim\limits_{t\rightarrow\infty}\mathbb{E}\left[\frac{1}{t}\sum_{\tau=0}^{t-1}\alpha(\tau)\right]\\
			\textrm{subject~to}~~&\limsup\limits_{t\rightarrow\infty} \mathbb{E}\left[\frac{1}{t}\sum_{\tau=0}^{t-1}\left(\alpha(\tau)-H(\tau)b_2(\tau)\right)\right]\leq 0, \label{condi2}\\
			& \overline{Q}< \infty, \\
			&0\leq  \alpha(t)\leq \alpha_{\max},\label{condi3} \\
			&\mathbf{U}(t)\in\{(1,1), (1,0), (0,1),(0,0)\}.
		\end{align}
	\end{subequations}
	The inequality condition in \eqref{condi2} can be transformed into a queue stability problem with the help of virtual queues. We define the virtual queue $Z(t)$ that updates by the following equation:
	\begin{equation}
		Z(t+1)=\max[Z(t)+\alpha(t)-H(t)b_2(t),0].
		\label{eq:evolution_Z}
	\end{equation}
	The rectangular constraint $0\leq  \alpha(t)\leq \alpha_{\max}$ is to make the auxiliary variable bounded, where $\alpha_{\max}$ is a sufficiently large constant.\footnote{Denote by $\alpha_{\text{opt}}$ the optimal solution to the stochastic optimization problem defined in \eqref{eq:opt-trans2} over all possible scheduling policies. When adding the rectangular constraint $0\leq  \alpha(t)\leq \alpha_{\max}$ to the problem,  $\alpha_{\max}$ should be chosen large enough to ensure $0\leq \alpha_{\text{opt}}\leq \alpha_{\max}$. More details can be found in \cite{neely2010stochastic}.} 
	Similar to the AoI optimization case, the DPP algorithm for the PAoI optimization problem is described in Algorithm \ref{algorithm:DSA}.
	\begin{algorithm}[t!]
		\caption{DPP Algorithm for PAoI Minimization} \label{algorithm:DSA}
		\begin{enumerate}
			\item Initialization: $Q(0)=0$, $B(0)=0$, $Z(0)=0$. Choose $\alpha_{\max}$ and $V$. Set $t=1$.
			\item At slot $t$, the network scheduler observes $Q(t)$, $Z(t)$ and $H(t)$, and makes decision $\mathbf{U}(t)$ by solving
			\begin{equation}
				\maximize\limits_{\mathbf{U}(t)}~~ Z(t)H(t)p_2(t)+Q(t)p_1(t),
				\label{eq:solution} \nonumber
			\end{equation} 
			where $p_1(t)$ and $p_2(t)$ are given in \eqref{eq:p1} and \eqref{eq:p2}.
			\item The auxiliary variable $\alpha(t)$ is chosen by
			\begin{equation}
				\alpha(t)= 
				\left\lbrace 
				\begin{array}{ccc}
					\alpha_{\max}
					& \text{if}~Z(t)\leq V, \\
					0
					& \text{otherwise}.
				\end{array} \right.\nonumber
			\end{equation}
			\item Update all queues by
			\begin{align} \notag
				Q(t+1)&=\max[Q(t)-b_1(t),0]+a_1(t),\\ \notag
				B(t+1)&=\max[B(t)-u_2(t),0]+a_2(t),\\ \notag
				Z(t+1)&=\max[Z(t)+\alpha(t)-H(t)b_2(t),0].
			\end{align}
			\item Repeat steps 2--4 for the next slot $t+1$.
		\end{enumerate}
	\end{algorithm}

	\begin{lemma}
		\label{lemma3}
		The DPP algorithms for both AoI and PAoI optimization problems guarantee that the data queue at node $S_1$ is strongly stable. The queue backlog $Q(t)$ of $S_1$ satisfies:
		\begin{equation}
			\limsup\limits_{t\rightarrow \infty} \frac{1}{t}\sum\limits_{\tau=0}^{t-1}\mathbb{E}[Q(\tau)]\leq \frac{C+V}{\epsilon}.
			\label{eq:backlog-bound}
		\end{equation}
		This inequality holds for any value of $\epsilon$ bounded by $0\leq \epsilon\leq \min\{p_{1/1}-\lambda, \delta\cdot  p_{2/1,2}\}$.
		Here, $C=\frac{\lambda^2+1}{2}$ for the case with AoI optimization and $C=\frac{\alpha_{\max}^2+\lambda^2+2 }{2}$ for the case with PAoI optimization. $V$ is a constant weight that affects the tradeoff between the performance optimization and the queue congestion.
	\end{lemma}
	\begin{IEEEproof}
		See Appendix \ref{appen3}.
	\end{IEEEproof}

	\subsection{Discussion on Further Extension}
	If there are multiple AoI-oriented and throughput-oriented nodes in the system, we can formulate an optimization problem with the weighted sum of the average AoI as the objective function and include as constraints the stability conditions for the throughput-oriented users. This problem can still be solved by using the standard Lyapunov optimization approach. The computational complexity of the DPP algorithm will depend on the scheduling decision space, which grows exponentially with $K$. However, in the PRA policy, since we have throughput-oriented users with bursty traffic, we face interaction among the queues. To characterize in closed-form the stable throughput in such setup is a well-known difficult problem for more than three users \cite{TWC2018}.
	
	\section{Simulation Results}
	\label{sec:simu}
	In this section, we evaluate the performance of the proposed PRA and DPP policies through simulations and compare the average AoI and PAoI obtained with these two policies.
	
	Since our analytical results do not depend on any specific channel model, the success probabilities we use in the simulations are divided into two cases: 1) strong MPR,
	$p_{1/1}=0.95$, $p_{1/1,2}=0.63$, $p_{2/2}=0.924$, $p_{2/1,2}=0.41$;  2) weak MPR, $p_{1/1}=0.924$, $p_{1/1,2}=0.515$, $p_{2/2}=0.882$, $p_{2/1,2}=0.3$.\footnote{The success probabilities are obtained by considering $\tilde{\beta}_1=12$\,dB for node $S_1$ and $\tilde{\beta}_1=10\,$dB for node $S_2$, with SNR/SINR threshold $-1$\,dB for the strong MPR case and $1$\,dB for the weak MPR case.}
	For the PRA policy, we choose $\xi=0.001$. For the DPP policy, we choose $V=200$ and $\alpha_{\max}=1$. We have chosen sufficiently long simulation time ($10^6$ slots) to make sure that the Markov chains reach their steady state behaviors.

	\begin{figure}[t!]
		\centering
		\includegraphics[width=\columnwidth]{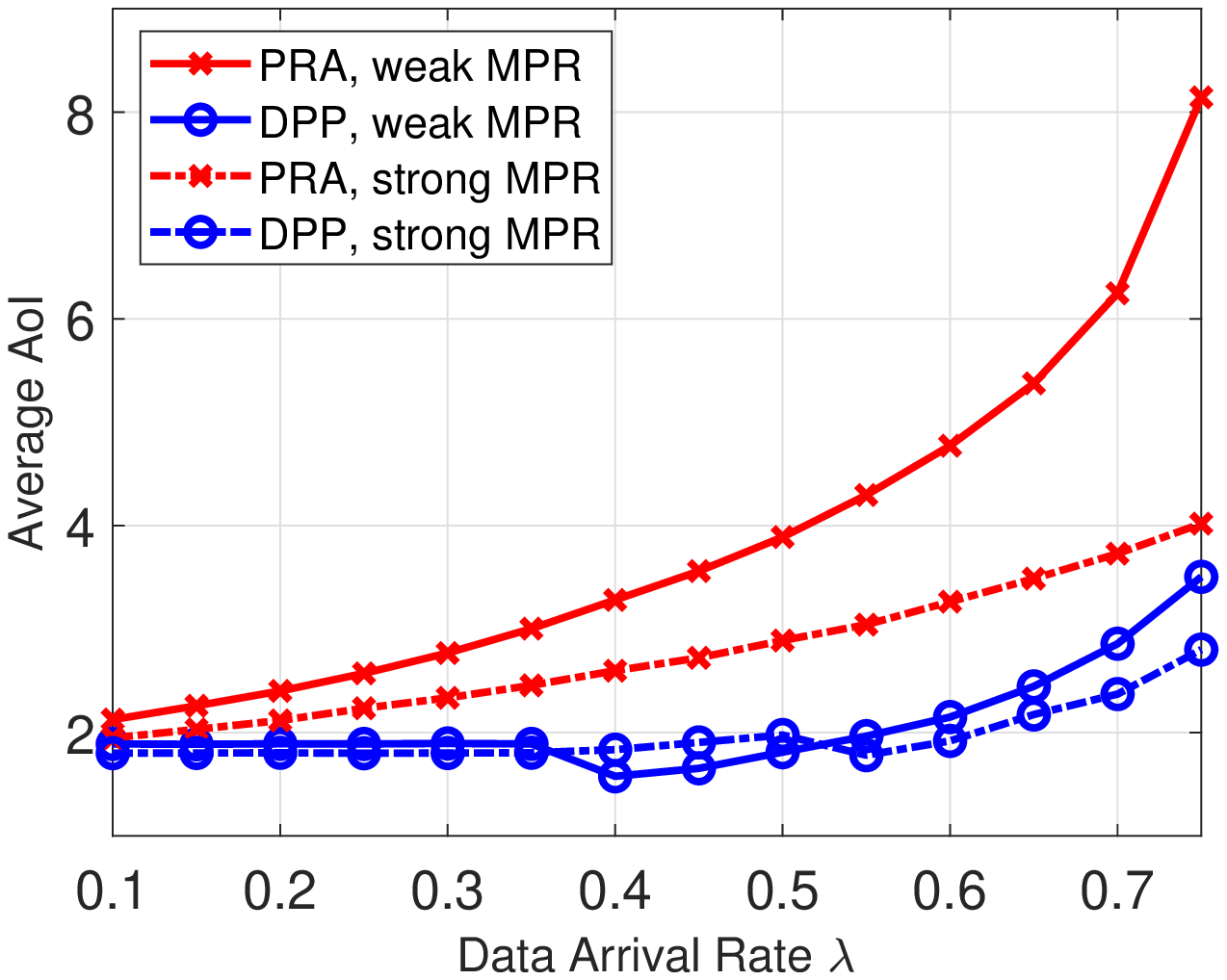}
		\caption{Average AoI vs. data arrival probability $\lambda$. Energy arrival probability $\delta=0.6$.}
		\label{fig:aoi_vs_lambda}
	\end{figure}  
	
	\begin{figure}
		\centering
		\begin{subfigure}[b]{0.5\textwidth}
			\centering
			\includegraphics[width=\columnwidth]{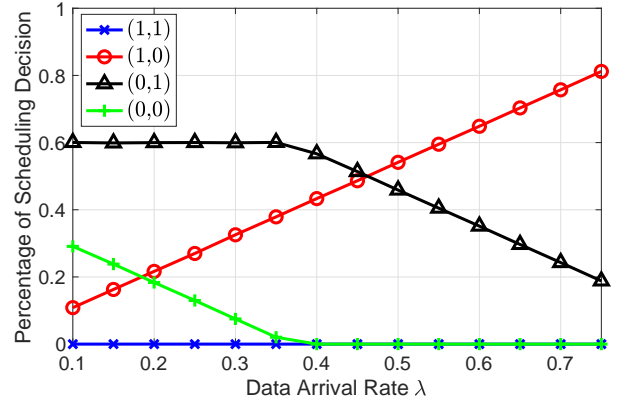}
			\label{fig:weak_percentage}
			\vspace{-0.3cm}
			\subcaption{Weak MPR}
		\end{subfigure}
		\hfill
		\begin{subfigure}[b]{0.5\textwidth}
			\centering
			\includegraphics[width=\columnwidth]{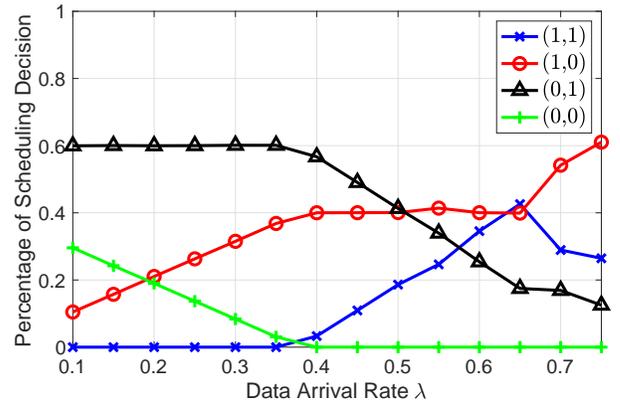}
			\label{fig:strong_percentage}
			\vspace{-0.3cm}
			\subcaption{Strong MPR}
		\end{subfigure}
		\caption{Percentage of time slots for each scheduling decision vs. data arrival probability $\lambda$, when the DPP policy is applied. Energy arrival probability $\delta=0.6$.}
		\label{fig:percen_vs_lambda}
	\end{figure}
	
	\begin{figure}[t!]
		\centering
		\includegraphics[width=\columnwidth]{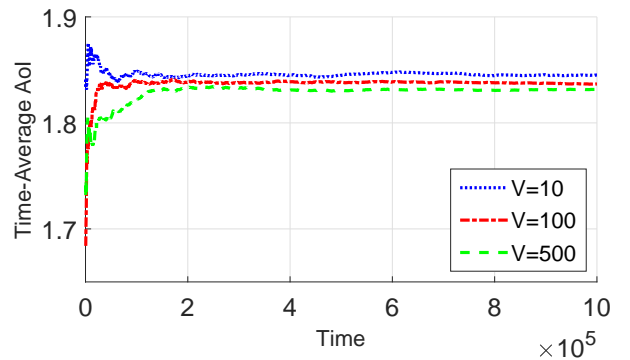}
		\caption{Time-average expected AoI vs. time for different values of $V$. Data arrival probability $\lambda=0.75$, energy arrival probability $\delta=0.6$. Strong MPR.}
		\label{fig:convergence_V}
	\end{figure}

	\subsection{Average AoI Comparison}
	In Fig.~\ref{fig:aoi_vs_lambda}, we compare the average AoI obtained with both policies, for different values of the data arrival probability $\lambda$.
	First, we observe that the DPP policy always achieves lower average AoI than the PRA policy. The difference is more significant when the data arrival probability is larger. 
	Second, the performance of the DPP policy is divided into two regimes. When $\lambda$ reaches a certain point, the average AoI has a sudden drop, and then increases with $\lambda$. To understand this phenomenon, in Fig.~\ref{fig:percen_vs_lambda} we present the percentage of each scheduling decision obtained with the DPP policy, when the destination node has weak and strong MPR capabilities, respectively. We see that in the case with weak MPR (in Fig.~\ref{fig:percen_vs_lambda}(a)), the two nodes are never active at the same time. However, in the strong MPR case as shown in Fig.~\ref{fig:percen_vs_lambda}(b), the probability of having two concurrent transmissions (the blue curve) starts to be non-zero after $\lambda$ reaches a certain point. One common observation from these two figures is that, when $\lambda\simeq0.4$, there is a turning point where the percentage of both nodes being idle becomes $0$, i.e., there are no idle slots in the channel. It means that the DPP policy makes full use of the transmission opportunities in the channel without wasting any slot.
	
	Note that the non-smoothness of the curves in  Fig.~\ref{fig:aoi_vs_lambda} and Fig.~\ref{fig:percen_vs_lambda}(b) is not because of insufficient simulation time, but comes from the performance-congestion tradeoff in the DPP policy. 
	Since the parameter $V$ determines how much weight we put on the penalty function, in Fig.~\ref{fig:convergence_V} we show the effect of $V$ on the time-average AoI. We see that larger $V$ gives smaller long-term average AoI, but the required time to reach the desired point also increases. For the value we choose, $10^{6}$ slots are sufficient for the DPP algorithm to converge.

	\begin{figure}[t!]
		\centering
		\begin{subfigure}[b]{0.5\textwidth}
			\includegraphics[width=\columnwidth]{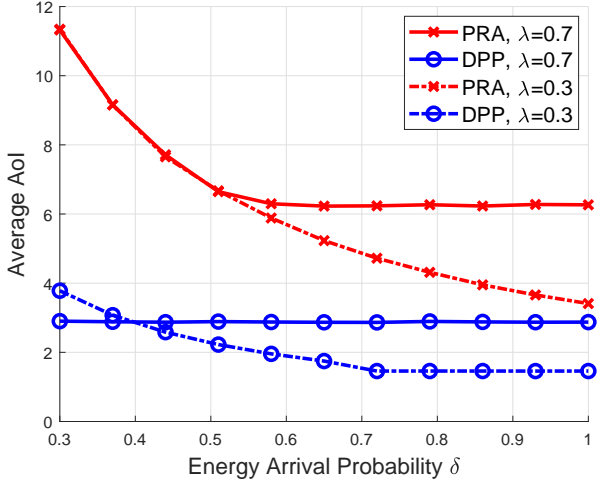}
			\label{fig:weak_delta}
			\vspace{-0.3cm}
			\subcaption{Weak MPR}
		\end{subfigure}
		\hfill
		\begin{subfigure}[b]{0.5\textwidth}
			\includegraphics[width=\columnwidth]{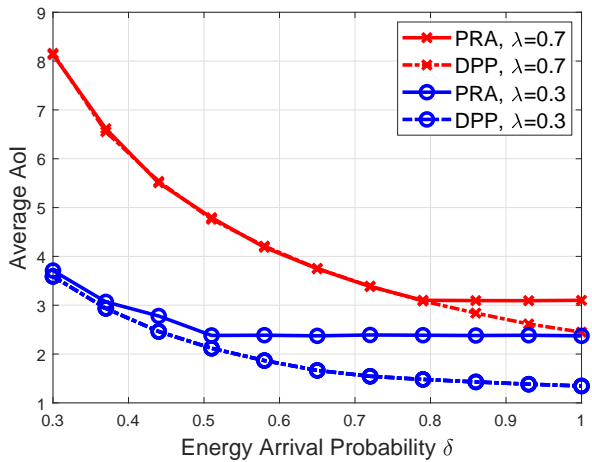}
			\label{fig:strong_delta}
			\vspace{-0.3cm}
			\subcaption{Strong MPR}
		\end{subfigure}
		\caption{Average AoI vs. energy arrival probability $\delta$. Data arrival probability $\lambda=\{0.3, 0.7\}$.}
		\label{fig:aoi_vs_delta}
	\end{figure}

	In Fig.~\ref{fig:aoi_vs_delta}, we show the relation between the average AoI and the energy arrival probability $\delta$. Also here, the DPP policy performs significantly better than the PRA policy, especially when the destination node has weak MPR capabilities. 
	Another interesting observation is, for both policies, with high data arrival probability, e.g., $\lambda=0.7$, the average AoI becomes independent of $\delta$ after $\delta$ reaches a certain threshold. This is because when $\lambda$ is large, $q_2$ should be small enough to satisfy the stability condition of node $S_1$. Then the age performance of $S_2$ is mainly limited by the energy departure process instead of by the arrivals.

	\begin{figure}[ht!]
		\centering
		\includegraphics[width=\columnwidth]{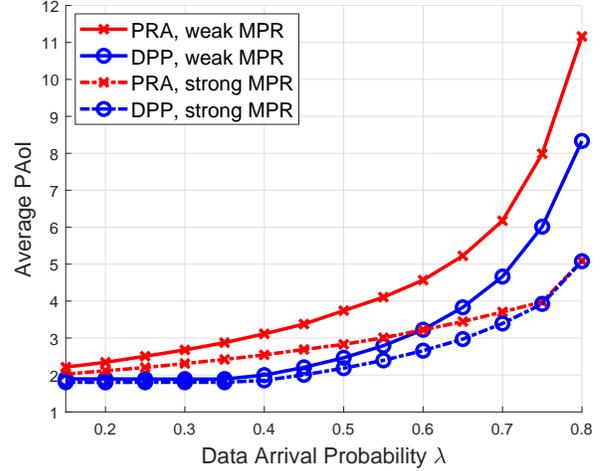}
		\caption{Average PAoI vs. data arrival probability $\lambda$. Energy arrival probability $\delta=0.6$.}
		\label{fig:paoi_vs_lambda}
	\end{figure} 
	
	Note that existing works such as \cite{Naware, TWC2018} have studied the stability region in two-user MAC with random access and MPR capabilities, which can provide a theoretical explanation for our findings from the stable throughput perspective. In these references, the weak/strong MPR capabilities correspond to $\frac{p_{1/1,2}}{p_{1/1}}+ \frac{p_{2/1,2}}{p_{2/2}}\lessgtr 1$. In the weak MPR case, the stable throughput of the random access MAC channel becomes a non-convex set, while it is a convex set for the strong MPR case. \textit{Convexity of the stable throughput region determines when a parallel concurrent transmission scheme is preferable to a time-sharing scheme.} Here, we observe that in the weak MPR case, the optimal strategy is to schedule one user at a time, which is not the case for the strong MPR.
	
	\subsection{Average PAoI Comparison}
	In Fig.~\ref{fig:paoi_vs_lambda}, we plot the average PAoI obtained with both policies as a function of the data arrival probability $\lambda$. First, we see that the DPP policy still outperforms the PRA policy in most cases. Second, with the DPP policy (the blue curves), when $\lambda$ increases, the average PAoI first remains the same, then increases with $\lambda$. This is because when $\lambda$ is small, the DPP algorithm will not allocate many transmission slots to node $S_1$. The EH node $S_2$ will have more chances for transmitting status updates when the battery is non-empty. Then the age of node $S_2$ is only limited by the energy arrivals. A third observation is that when $\lambda$ is high, with strong MPR, the performances of these two policies become the same, which is different than what we have observed in Fig.~\ref{fig:aoi_vs_lambda}.

	\begin{figure}[h!]
		\centering
		\begin{subfigure}[b]{0.5\textwidth}
			\includegraphics[width=\columnwidth]{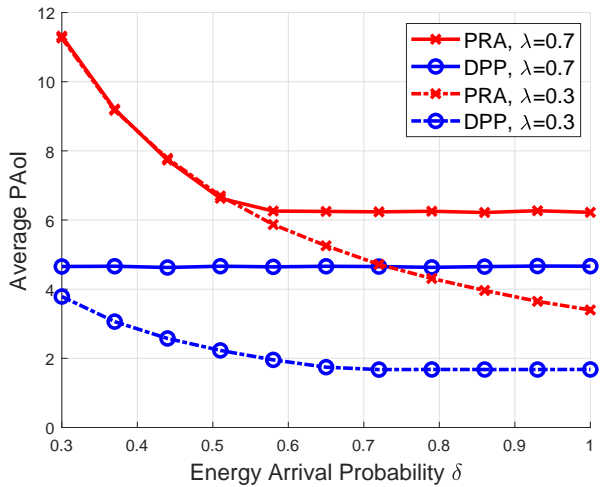}
			\label{fig:paoi_delta_weak}
			\vspace{-0.3cm}
			\subcaption{Weak MPR}
		\end{subfigure}
		\hfill
		\begin{subfigure}[b]{0.5\textwidth}
			\includegraphics[width=\columnwidth]{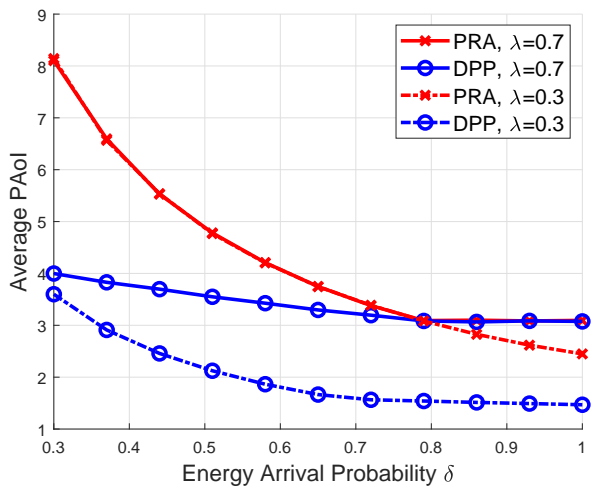}
			\label{fig:paoi_delta_strong}
			\vspace{-0.3cm}
			\subcaption{Strong MPR}
		\end{subfigure}
		\caption{Average PAoI vs. energy arrival probability $\delta$. }
		\label{fig:paoi_delta}
	\end{figure}
	
	In Fig.~\ref{fig:paoi_delta}, we present the average PAoI obtained with both policies for different values of the data and energy arrival probabilities. Similar to the previous results, we observe the advantage of the DPP policy, especially in the case with low MPR capabilities. The only exceptional case when the PRA policy performs as good as the DPP policy is when both $\lambda$ and $\delta$ are very high, and with strong MPR capabilities at the destination node, as shown in Fig.~\ref{fig:paoi_delta}(b). Recall that for both policies, PAoI-optimal scheduling is also throughput-optimal. \textit{This observation suggests that with strong MPR and high data and energy arrival probabilities, the two policies achieve the same optimal throughput.} 
	
	Comparing the results in Fig.~\ref{fig:aoi_vs_delta} and in Fig.~\ref{fig:paoi_delta}, we see that when $\lambda$ is small, e.g., $\lambda=0.3$, the average AoI and PAoI achieved with the DPP policy are very close. However, when $\lambda$ is large, e.g., $\lambda=0.7$, the average PAoI obtained with the DPP policy is obviously higher than the average AoI. This observation shows that with the DPP policy, PAoI-optimal scheduling is not equivalent to AoI-optimal in the general case.

	\section{Conclusions}
	We studied age-optimal scheduling in a MAC with two heterogeneous source nodes generating different types of data traffic. One grid-connected node has bursty data arrivals and another node with EH capabilities sends fresh status updates to a common destination. An optimization problem was formulated to minimize the average age of the EH node with respect to the queue stability of the grid-connected node. We solved this problem by considering a PRA policy with independent and random transmission decisions, and a DPP policy developed by using Lyapunov optimization. We derived the optimal solution with the PRA policy in closed form, and proved that the average AoI is the same as the average PAoI. For the DPP policy, we proposed two dynamic scheduling algorithms for the AoI and PAoI optimization problems, and showed that the PAoI-optimal scheduling is also throughput-optimal, but not AoI-optimal. Simulation results showed that the DPP policy significantly outperforms the PRA policy, especially when the destination node has weak MPR capabilities. 
	
	\appendix
	\appendices
	
	\subsection{Proof of Theorem \ref{theorem1}}
	\label{appen_theorem}
	Between two successful receptions of status updates, there might be more than one attempted transmission. Let $T_i$ represent the time difference between the $(i-1)$-th and the $i$-th attempted transmissions, from Fig.~\ref{fig:aoi}, we have
	\begin{equation}
		X=\sum_{i=1}^{M}T_i,
		\label{eq:X_k}
	\end{equation} 
	where $M$ is a RV representing the number of attempted transmissions between two successfully received status updates. Note that when $i=1$, $T_1$ represents the time difference between the latest successfully received update and the first attempted transmission after that. Since $T_i$ is a stationary stochastic process, in the following we use $\mathbb{E}[T]$ to denote the expected value of $T_i$ for an arbitrary $i$. The expected values of $X$ can be obtained by using the probability mass function of a geometric distribution, which gives
	\begin{equation} \label{EX-gen}
		\mathbb{E}[X]=\sum\limits_{m=1}^{\infty}m \mathbb{E}[T](1-\overline{p}_2)^{m-1}\overline{p}_2=\frac{\mathbb{E}[T]}{\overline{p}_2},
	\end{equation}
	where $\overline{p}_2$ is the success probability of the transmission from $S_2$, which is the weighted sum of $p_{2/2}$ and $p_{2/1,2}$. When the queue at $S_1$ is stable, we have
	\begin{align}
		\overline{p}_2=&~p_{2/2}\left(1-q_1\cdot \mathbb{P}[Q(t)\neq 0]\right)+p_{2/1,2} q_1 \mathbb{P}[Q(t)\neq 0]\nonumber\\
		=&~p_{2/2}- \frac{(p_{2/2}-p_{2/1,2}) \lambda}{p_{1/1}-\min\{\delta, q_2\}\cdot \left(p_{1/1}-p_{1/1,2}\right)}.
		\label{eq:pqne}
	\end{align}
	
	For the second moment of $X$, we start from
	\begin{equation}
		X^2=\left(\sum\limits_{i=1}^{M}T_i\right)^2=\sum\limits_{i=1}^{M}T_i^2+\sum\limits_{i=1}^{M}\sum\limits_{j=1,j\neq i}^{M}T_i T_j.
	\end{equation}
	Taking conditional expectation of both sides, we get
	\begin{equation}
		\mathbb{E}[X^2 \vert M]=M\mathbb{E}[T^2]+M(M-1)\left(\mathbb{E}[T]\right)^2.
	\end{equation}
	Then we have
	\begin{align} \label{EX2-gen}
		\mathbb{E}[X^2 ]&=\sum\limits_{m=1}^{\infty}\mathbb{E}[X^2 \vert m] (1-\overline{p}_2)^{m-1}\overline{p}_2\nonumber\\
		&=\frac{\mathbb{E}[T^2 ]}{\overline{p}_2}+\left(\mathbb{E}[T]\right)^2\frac{2(1-\overline{p}_2)}{\overline{p}_2^2}.
	\end{align}
	Here, the sum converges when $\overline{p}_2>0$. Substituting \eqref{EX-gen} and \eqref{EX2-gen} into \eqref{eq:aoi}, the average AoI becomes
	\begin{equation} 
		\overline{A} = \frac{\mathbb{E}[T^2 ]}{2\mathbb{E}[T]} + \frac{\mathbb{E}[T](1-\overline{p}_2)}{\overline{p}_2} + \frac{1}{2}.\label{eq:average_aoi}
	\end{equation}
	
	Since $T$ is a discrete number representing the time difference between two consecutive attempted transmissions, the probability mass function of $T$ is given by
	\begin{align}
		\mathbb{P}[T=k]=~&\mathbb{P}[B(t)=0]\sum\limits_{l=1}^{k-1}(1-\delta)^{l-1}\delta (1-q_2)^{k-l-1}q_2\nonumber\\
		&+\mathbb{P}[B(t)\neq 0](1-q_2)^{k-1}q_2.
		\label{eq:T}
	\end{align}
	The first term is the probability that when the battery of $S_2$ is empty, $S_2$ does not attempt to transmit for $k-1$ consecutive slots, either because of no energy arrival or because it decides not to transmit. The second term is the probability that when the battery is non-empty, $S_2$ decides not to transmit for $k-1$ consecutive slots.
	
	Since \eqref{eq:T} involves $\mathbb{P}[B(t)=0]$, as mentioned previously, our analysis will be given in two cases: when $\delta<q_2$ and when $\delta=1$, respectively.
	
	\subsubsection{When $S_2$ relies on EH and $\delta<q_2$} \label{sec:S2EH}
	Recall that we have $\mathbb{P}[B(t)\neq 0]=\frac{\delta}{q_2}$ when $\delta<q_2$.
	After substituting $\mathbb{P}[B(t)\neq 0]=\frac{\delta}{q_2}$ into \eqref{eq:T}, we have
	\vspace{-0.2cm}
	\begin{align}
		\vspace{-0.2cm}
		P(T=k) =& \left(1-\frac{\delta}{q_2}\right)q_2(1-q_2)^{k-2}\delta\sum\limits_{l=1}^{k-1}\left(\frac{1-\delta}{1-q_2}\right)^{l-1}\nonumber\\
		&+\frac{\delta}{q_2}(1-q_2)^{k-1}q_2  \nonumber\\
		=&\left(q_2-\delta\right)(1-q_2)^{k-2}\delta\sum\limits_{l=1}^{k-1}\left(\frac{1-\delta}{1-q_2}\right)^{l-1}\nonumber\\&+\delta(1-q_2)^{k-1}.
	\end{align}
	After some simplification, we can obtain
	\vspace{-0.2cm}
	\begin{equation}
		\mathbb{E}[T]=\sum\limits_{k=1}^{\infty}k\mathbb{P}[T=k]=\frac{1}{\delta}.
		\label{eq:ET}
	\end{equation}
	Similarly, we have
	\vspace{-0.2cm}
	\begin{equation}
		\mathbb{E}[T^2]=\sum\limits_{k=1}^{\infty}k^2 \mathbb{P}[T=k] = \frac{2-\delta}{\delta^2}.
		\label{eq:ETsqure}
	\end{equation}
	
	\subsubsection{When $S_2$ is connected to a power grid}
	In this case, we have $\mathbb{P}[B(t)\neq 0]=1$. From \eqref{eq:T} we get
	\begin{eqnarray}
		\vspace{-0.2cm}
		\mathbb{P}[T=k]&=&(1-q_2)^{k-1}q_2, \\
		\mathbb{E}[T]&=&\sum\limits_{k=1}^{\infty}k\mathbb{P}[T=k]=\frac{1}{q_2}, \label{eq:ET_connected}\\
		\mathbb{E}[T^2]&=&\sum\limits_{k=1}^{\infty}k^2 \mathbb{P}[T=k]=\frac{2-q_2 }{q_2^2}.\label{eq:ET2_connected}
	\end{eqnarray}
	
	Combining these two cases, we have that 
	\begin{eqnarray}
		\vspace{-0.2cm}
		\mathbb{E}[T]&=&\frac{1}{\min\{q_2,\delta\}}, \label{eq:ET_both}\\
		\mathbb{E}[T^2]&=&\frac{2-\min\{q_2,\delta\} }{\min\{q_2,\delta\}^2}.  \label{eq:ET2_both}
	\end{eqnarray}
	Substituting \eqref{eq:ET_both} and \eqref{eq:ET2_both} into \eqref{eq:average_aoi}, we obtain Theorem~\ref{theorem1}.
	
	\subsection{Proof of Lemma \ref{lemma_optimal_proba}}
	\label{appen1}
	When $S_2$ relies on EH and assuming $\delta<q_2$, the average AoI is $1/(\delta \cdot \overline{p}_2)$, which is independent of $q_1$ and $q_2$ as it can be seen from \eqref{eq:pqne}. Thus, the optimal value of $q_2$ can be any value within the range $(\delta, 1]$.
	From the queue stability condition given in \eqref{eq:stability_q1}, we have 
	\begin{equation}
		q_1>\frac{\lambda}{p_{1/1}(1-\delta)+\delta p_{1/1,2}}. \label{eq:q1star}
	\end{equation}
	Since the probability $q_1$ cannot be larger than $1$, we have $\lambda<p_{1/1}(1-\delta)+\delta p_{1/1,2}$, which corresponds to $\delta<\frac{p_{1/1}-\lambda}{p_{1/1}-p_{1/1,2}}$. Combining with the condition $0\leq \delta\leq 1$, we have
	\begin{equation}
		0<\delta< \min\left\{\frac{p_{1/1}-\lambda}{p_{1/1}-p_{1/1,2}},1\right\}.
	\end{equation}
	If this condition is satisfied, the optimal transmit probabilities are $q_1^*>\frac{\lambda}{p_{1/1}(1-\delta)+\delta p_{1/1,2}}$ and $q_2^*>\delta$.
	
	If $\delta
	\geq \min\left\{\frac{p_{1/1}-\lambda}{p_{1/1}-p_{1/1,2}},1\right\}$, the optimal values of $q_1$ cannot be found by \eqref{eq:q1star} because the threshold exceeds $1$. Thus, the queue stability implies that $\delta\geq q_2$. In this case, we can disregard the energy queue and consider the system as if $S_2$ was connected to a power grid. Then the AoI optimization problem becomes
	\begin{subequations}
		\begin{align}
			\minimize\limits_{ q_1,q_2} ~~&\frac{1}{q_2 \overline{p}_2} \\
			\textnormal{subject~to}~~& q_1 (1-q_2)p_{1/1}+q_1 q_2p_{1/1,2}>\lambda, \label{eq:stability}\\
			& 0\leq q_1\leq 1,\\
			& 0\leq q_2 \leq  \delta.
		\end{align}
		\label{eq:optim_connectted}%
	\end{subequations}
	Since the average AoI is inversely proportional to $q_2$, the optimal value of $q_2$ is the maximum value of $q_2\in[0,\delta]$ that satisfies the queue stability condition in \eqref{eq:stability}. Then we have
	\begin{equation}
		q_2<\frac{p_{1/1}-\lambda/q_1}{p_{1/1}-p_{1/1,2}}.
	\end{equation}
	The maximum value of $q_2$ is achieved when $q_1=1$. Then we obtain the optimal solution as $q_1^*=1$ and $q_2^*=\min\left\{\frac{p_{1/1}-\lambda}{p_{1/1}-p_{1/1,2}},\delta\right\}-\xi$, where $\xi>0$ is a small positive value to ensure that the service probability is strictly larger than the arrival probability.
	
	\subsection{Proof of Lemma \ref{lemma2}}
	\label{appen2}
	From the queue evolution $Q(t+1)=\max[Q(t)-b_1(t),0]+a_1(t)$,
	the Lyapunov drift is bounded by
	\begin{equation}
		\begin{split}
			\Delta(t)\leq&~ \frac{1}{2}\mathbb{E}\left[a_1(t)^2+b_1(t)^2|\boldsymbol{\Phi}(t)\right]\\&+\mathbb{E}[Q(t)\left(a_1(t)-b_1(t)\right)\!|\boldsymbol{\Phi}(t)].
		\end{split}
	\end{equation}
	From $\mathbb{E}[a_1(t)|\boldsymbol{\Phi}(t)]=\mathbb{E}[a_1(t)]=\lambda$ and $\mathbb{E}[b_1(t)^2|\boldsymbol{\Phi}(t)]\leq 1$, we have
	\begin{equation}
		\Delta(t)\leq \frac{\lambda^2+1}{2}+\lambda Q(t)-\mathbb{E}[Q(t)b_1(t)|\boldsymbol{\Phi}(t)].
	\end{equation}
	Adding the penalty term $V \mathbb{E}[A(t+1) | \boldsymbol{\Phi}(t)]$, with $A(t+1)=A(t)+1-H(t)b_2(t)A(t)$ and $p_i(t)=\mathbb{E}[b_i(t)]$ for $i=\{1,2\}$, we obtain the upper bound for the DPP as in \eqref{eq:dpp-bound}.

	\subsection{Proof of Lemma \ref{lemma3}}
	\label{appen3}
	\subsubsection{Proof for queue stability with PAoI optimization}
	\label{appen2-part1}
	
	We define $\boldsymbol{\Theta}(t)=[Z(t),Q(t)]$ and consider the Lyapunov function $\mathcal{L}(\boldsymbol{\Theta}(t))=\frac{1}{2}Q(t)^2+\frac{1}{2}Z(t)^2$. The one-slot conditional Lyapunov drift is given by
	\begin{equation}
		\Delta(t)=\mathbb{E}[\mathcal{L}(\boldsymbol{\Theta}(t+1))-\mathcal{L}(\boldsymbol{\Theta}(t))|\boldsymbol{\Theta}(t)].
	\end{equation} 
	From \eqref{eq:queue-update} and \eqref{eq:evolution_Z},
	the Lyapunov drift is bounded by
	\begin{equation}
		\begin{split}
			\Delta(t)\leq &~\frac{\mathbb{E}[\alpha(t)^2+\!H(t)^2b_2(t)^2\!+\!a_1(t)^2\!+\!b_1(t)^2|\boldsymbol{\Theta}(t)]}{2}\\&+\mathbb{E}[Q(t)\left(a_1(t)-b_1(t)\right)|\boldsymbol{\Theta}(t)] \\&+\mathbb{E}[Z(t)\left(\alpha(t)-H(t)b_2(t)\right)|\boldsymbol{\Theta}(t)].
		\end{split}
	\end{equation}
	We know that $\mathbb{E}[a_1(t)|\boldsymbol{\Theta}(t)]=\lambda$, $\mathbb{E}[\alpha(t)|\boldsymbol{\Theta}(t)]\leq \alpha_{\max}$, $\mathbb{E}[H(t)^2b_2(t)^2|\boldsymbol{\Theta}(t)]\leq 1$, and\\ $\mathbb{E}[b_1(t)^2|\boldsymbol{\Theta}(t)]\leq 1$. Define $C=\frac{\alpha_{\max}^2+\lambda^2+2 }{2}$, we have
	\begin{equation}
		\begin{split}
			\Delta(t)\leq &~C+\lambda Q(t)-\mathbb{E}[Q(t)b_1(t)|\boldsymbol{\Theta}(t)] \\&+\mathbb{E}[Z(t)\left(\alpha(t)-H(t)b_2(t)\right)|\boldsymbol{\Theta}(t)].
		\end{split}
	\end{equation}
	Since $p_i(t)=\mathbb{E}[b_i(t)]$ for $i=\{1,2\}$,
	the drift-plus-penalty is bounded by
	\begin{equation}
		\begin{split}
			\Delta(t)\!-\!V \mathbb{E}\left[\alpha(t) |\boldsymbol{\Theta}(t)\right]\!\leq\!& ~C\!+\!\lambda Q(t)\!+\!\mathbb{E}[\alpha(t)(Z(t)-V)|\boldsymbol{\Theta}(t)]\\& \!+\!\mathbb{E}[Z(t)H(t)p_2(t)\!+\!Q(t)p_1(t)|\boldsymbol{\Theta}(t)],
		\end{split}
		\label{eq:dpp}
	\end{equation}
	where $V$ denotes the weight on the penalty function.

	We consider opportunistically minimizing the conditional expectation of the upper bound on the DPP, which results in the following two sub-problems:
	\begin{itemize}
		\item Observe $Z(t)$, $H(t)$ and $Q(t)$, choose the scheduling decision that maximizes $Z(t)H(t)p_2(t)+Q(t)p_1(t)$;
		\item Choose $0\leq \alpha(t)\leq \alpha_{\max}$ that minimizes $\alpha(t)(Z(t)-V)$, which gives
		\begin{equation}
			\alpha(t)= 
			\left\lbrace 
			\begin{array}{ccc}
				\alpha_{\max}
				& \text{if}~Z(t)\leq V, \\
				0
				& \text{otherwise}.
			\end{array} \right.\nonumber
		\end{equation}
	\end{itemize}
	Then we obtain the DPP algorithm described in Algorithm \ref{algorithm:DSA}.

	In the following, we show that this algorithm stabilizes the network. Consider an alternative S-only policy that makes stationary randomized decisions independent of the queue backlog, there exists $\epsilon>0$ such that the resulting values $p_1^{*}(t)$, $p_2^{*}(t)$, $H^{*}(t)$ and $\alpha^{*}(t)$ of the S-only policy satisfy:
	\begin{eqnarray}
		\mathbb{E}[p_1^{*}(t)]=p_1^{*}&\geq& \lambda+\epsilon, \label{ineq1}\\
		\mathbb{E}[H^{*}(t)p_2^{*}(t)]-\mathbb{E}[\alpha^{*}(t)] &\geq&\epsilon,\label{ineq2}\\
		\mathbb{E}[\alpha^{*}(t)]&=  &\alpha_{sr}(\epsilon).\label{ineq3}
	\end{eqnarray}
	Here, $0\leq \alpha_{sr}(\epsilon)\leq \alpha_{\text{opt}}$ is a feasible solution to the problem defined in \eqref{eq:opt-trans2} that can be achieved by an S-only policy. The conditions in \eqref{ineq1} and \eqref{ineq2} are feasible for any $\epsilon$ bounded by $0\leq \epsilon\leq \min\{p_{1/1}-\lambda, \delta\cdot p_{2/1,2}\}$, if $\lambda<p_{1/1}$ holds.
	
	Since our DPP algorithm minimizes the right-hand side of \eqref{eq:dpp} in every slot, after taking iterated expectations on both sides, any alternative policy (S-only or not) would results in larger value for the expectation of the right-hand side of \eqref{eq:dpp}. Then we have
	\begin{equation}
		\begin{split}
			&\mathbb{E}[\mathcal{L}(\boldsymbol{\Theta}(t+1))]-\mathbb{E}[\mathcal{L}(\boldsymbol{\Theta}(t))]-V \mathbb{E}\left[\alpha(t) \right]\\ &~~\leq C-\epsilon \mathbb{E}[Q(t)]-\epsilon \mathbb{E}[Z(t)]- V\alpha_{sr}(\epsilon).
		\end{split}
		\label{eq:queue-size}
	\end{equation}
	Summing over $\tau=0,\ldots, t-1$, after rearranging terms, we obtain
	\begin{equation}
		\begin{split}
			\sum\limits_{\tau=0}^{t-1}\mathbb{E}[Q(\tau)]+\sum\limits_{\tau=0}^{t-1}\mathbb{E}[Z(\tau)]\leq t\frac{C+V(\mathbb{E}[\alpha(t)]-\alpha_{sr}(\epsilon))}{\epsilon}\\-\frac{\mathbb{E}[\mathcal{L}(\boldsymbol{\Theta}(t))]-\mathbb{E}[\mathcal{L}(\boldsymbol{\Theta}(0))]}{\epsilon}.
		\end{split}
	\end{equation}
	After neglecting the non-negative terms and dividing both sides by $t$, we get 
	\vspace{-0.1cm}
	\begin{equation}
		\begin{split}
			\frac{1}{t}\sum\limits_{\tau=0}^{t-1}\mathbb{E}[Q(\tau)]\leq \frac{C+V(\mathbb{E}[\alpha(t)]-\alpha_{sr}(\epsilon))}{\epsilon}\\+\frac{\mathbb{E}[\mathcal{L}(\boldsymbol{\Theta}(0))]}{t\epsilon}.
		\end{split}
	\end{equation}
	Since $\mathbb{E}[\alpha(t)]\leq \alpha_{\text{opt}}$ where $\alpha_{\text{opt}}$ is the optimal solution over all scheduling policies, taking the limit $t\rightarrow \infty$, we obtain
	\vspace{-0.1cm}
	\begin{equation}
		\limsup\limits_{t\rightarrow \infty} \frac{1}{t}\sum\limits_{\tau=0}^{t-1}\mathbb{E}[Q(\tau)]\leq \frac{C+V(\alpha_{\text{opt}}-\alpha_{sr}(\epsilon))}{\epsilon}.
	\end{equation}
	This shows that the queue of $S_1$ is strongly stable. 
	Furthermore, knowing that $0\leq \alpha_{\text{opt}}< 1$ because $\mathbb{E}[H(t)b_2(t)]\leq 1$, and $0\leq \alpha_{sr}(\epsilon)\leq \alpha_{\text{opt}}$, the queue backlog is bounded by
	\begin{equation}
		\limsup\limits_{t\rightarrow \infty} \frac{1}{t}\sum\limits_{\tau=0}^{t-1}\mathbb{E}[Q(\tau)]\leq \frac{C+V}{\epsilon}.
	\end{equation}
	This inequality holds for any value of $\epsilon$ bounded by $0\leq \epsilon\leq \min\{p_{1/1}-\lambda, \delta\cdot  p_{2/1,2}\}$.
	
	\subsubsection{Proof for queue stability with AoI optimization}
	Same as in the PAoI optimization case, we consider an alternative S-only policy that satisfies $\mathbb{E}[p_1^{*}(t)]\geq \lambda+\epsilon$. From the DPP bound in \eqref{eq:dpp-bound}, following similar steps as in Appendix \ref{appen2-part1}, we have
	\begin{align}
		&~\mathbb{E}[\mathcal{L}(\boldsymbol{\Phi}(t+1))]-\mathbb{E}[\mathcal{L}(\boldsymbol{\Phi}(t))]+V \mathbb{E}\left[H_2(t)p_2(t)A(t)\right]\nonumber\\~\leq &~ C-\epsilon \mathbb{E}[Q(t)]+V\mathbb{E}[A(t)+1-A(t+1)].
	\end{align}
	After neglecting some non-negative terms, we obtain
	\begin{align}
		\epsilon \mathbb{E}[Q(t)]\leq &~C+V+V\mathbb{E}[A(t)-A(t+1)]+\mathbb{E}[\mathcal{L}(\boldsymbol{\Phi}(t))]\nonumber\\&-\mathbb{E}[\mathcal{L}(\boldsymbol{\Phi}(t+1))].
		\label{eq:bound1}
	\end{align}
	Summing over $\tau=0,\ldots, t-1$, dividing both sides by $t\cdot \epsilon$,  and taking the limit $t\rightarrow\infty$, we obtain the same inequality as in \eqref{eq:backlog-bound}.

\end{document}